\documentclass[apssuperscriptaddress,amsmath,singlecolumn,nofootinbib,floatfix,nobalancelastpage]{revtex4-1}
\usepackage{graphicx}
\usepackage{latexsym}
\usepackage{amssymb}
\usepackage{amsmath}
\usepackage{subfig}
\usepackage{graphicx}
\usepackage{amsfonts}
\usepackage{bm}
\usepackage{multirow}
\usepackage{float}
\usepackage{color}

\newcommand{\beq}{\begin{equation}}
\newcommand{\eeq}{\end{equation}}
\newcommand{\beqn}{\begin{eqnarray}}
\newcommand{\eeqn}{\end{eqnarray}}
\usepackage{amsmath,amssymb,amsfonts,graphics,multirow}
\begin{document}

\def\ppnumber{\vbox{\baselineskip14pt
}}

\def\ppdate{
} \date{\today}

\title{\bf Quantum Liquids: Emergent higher-rank gauge theory and fractons}
\author{Yizhi You}
\affiliation{Department of Physics, Northeastern University, Boston, MA, 02115}

\begin{abstract}
Fracton emerges from strongly interacting many-body systems whose excitations, referred to as sub-dimensional particles, have restricted mobility or kinetic motions. These entities have garnered significant interest due to their interdisciplinary implications spanning topological quantum codes, quantum field theory, emergent gravity, quantum information, and more, revealing unique nonequilibrium behaviors such as nonergodicity and glassy dynamics. This review presents a structured and educational overview of fracton phenomena, specifically focusing on gapless fracton liquids. Noteworthy for their compressibility and gapless excitations, fracton liquids facilitate collective modes reminiscent of those gauge fluctuations found in Maxwell's electromagnetic framework. However, they are distinct due to an additional higher-moment conservation law that restricts the mobility of individual charges and monopoles.
Our exploration begins with the theoretical foundation of 3D fracton liquids, presenting a variety of emergent symmetric tensor gauge theories and reviewing their equilibrium and dynamic properties. Following this theoretical groundwork, we discuss the material realization of various fracton liquids in Yb-based breathing pyrochlore lattices, close-packed tiling structures and other synthetic quantum matter platforms.
Additionally, we introduce a general protocol to foster and manipulate emergent fracton spin liquids in the realm of frustrated magnetism.

 Keywords: fracton, higher-rank gauge theory, tensor electromagnetism, pinch point, close-packed tiling system
\end{abstract}

\maketitle

\bigskip
\newpage

\section{Prelude: Fracton and higher-rank gauge theories}

Understanding the nature of symmetry and constraint in many-body theory has led to significant developments in the theoretical exploration of emergent phases with long-range entanglement. These developments have potential applications ranging from designing correlated materials to creating quantum devices. Recently, there has been growing interest in a new class of quantum phases known as the \textit{Fracton} phase, characterized by a rich interplay between correlations, symmetry, topology, and dynamics. Fractons emerge from strongly interacting many-body systems whose excitations, referred to as sub-dimensional particles, have restricted mobility or kinetic motions\cite{Vijay2015-jj,Haah2011-ny,Chamon2005-fc,hermele17,bravyi2010topological,nandkishore2019fractons,pretko2020fracton,slagle2017fracton,gromov2024colloquium}. These unusual particles were first encountered in quantum spin models\cite{Vijay2015-jj,Haah2011-ny,Chamon2005-fc,hermele17,bravyi2010topological} but have since been shown to arise in various contexts, including topological crystalline defects\cite{pretko2017fracton,pretko2019crystal,nguyen2020fracton}, vortex hydrodynamics\cite{doshi2021vortices,du2024noncommutative,pretko2018symmetry}, and frustrated paramagnets\cite{yan2019rank,han2022realization,Halasz2017-ov}. Although certain classes of gapped fracton phases share some features with conventional topological phases, such as locally indistinguishable ground states with long-range entanglement\cite{Vijay2015-jj,Vijay2016-dr,Chamon2005-fc,Haah2011-ny,yoshida2013exotic,pretko2020fracton} — they are explicitly distinct from the usual topologically ordered phases. Their physical features depend on the geometry of the system and are sensitive to the UV cutoff\cite{slagle2019symmetric,you2019emergent,xu2007bond,paramekanti2002ring,tay2010possible,seiberg2020exotic,gorantla2021low,ye2022ultraviolet,zhou2021fractal,delfino20232d}. These exotic phases extend and challenge our existing notions of topological field theory\cite{ shirley2019universal,shirley2017fracton,shirley2018fractional,slagle2019symmetric,slagle2021foliated,yoshida2015bosonic} and the renormalization group\cite{seiberg2020exotic,lake2021rg,xu2007bond,paramekanti2002ring,haah2014bifurcation,ma2018topological,he2018entanglement,shirley2019universal,you2021fractonic}. They have attracted broad interdisciplinary interest as they lie at the intersection of multiple fields, including topological physics, quantum field theory, emergent gravity, quantum information, glassy dynamics, and liquid crystals. Furthermore, the restricted mobility of fractons causes them to exhibit unusual nonequilibrium properties\cite{newman1999glassy,pretko2017fracton,seiberg2020exotic,pretko2017finite,pai2018fractonic}, such as nonergodic behavior and glassy dynamics\cite{prem2017glassy,sala2020ergodicity,pai2019localization,feldmeier2020emergent,pai2019dynamical}, which provides a feasible platform for stable quantum memory in which information can be stored in a way that is intrinsically robust to random noise.

The first manifestation of fracton quasiparticles was encountered in a class of exactly solvable gapped spin models\cite{Vijay2015-jj,Vijay2016-dr,Chamon2005-fc,Haah2011-ny,yoshida2013exotic,williamson2016fractal}, known as fracton topological phases. These phases are characterized by the presence of topologically non-trivial excitations that can only move on lower-dimensional submanifolds. Many fracton phases exhibit behaviors that, in some respects, are similar to topological order. They feature robust ground-state degeneracies and statistical interactions between point-like quasiparticles. However, they differ qualitatively from topologically ordered phases in several aspects. The ground state degeneracy is not topological but rather sensitive to geometric factors such as system sizes and twisted boundary conditions. Excitations are generally sub-dimensional, meaning they are either immobile or their motion is restricted to lines, planes, or fractals. This phenomenon challenges our understanding of topological order, as the low-energy theory, including its ground state degeneracy on closed manifolds, depends on some non-topological features of space. Moreover, changing the geometry's curvature produces a qualitative change in the kinetic motion of anyon excitations and, at the same time, alters the ground state degeneracy on closed manifolds\cite{slagle2018symmetric,yan2022cube,manoj2020screw,you2019non,shirley2018foliated}.

Alternatively, a majority of fracton phases can be viewed \textit{higher-rank gauge theories}\footnote{Also dubbed `symmetric tensor gauge theory' in some literature.} obtained by gauging the \textit{charge multipole algebra}
---a generalization of space-symmetries of a charged matter that displays conservation of various components of the multipole moments in the charge density
\cite{bulmash2018higgs,pretko2018fracton,pretko2017generalized,pretko2017generalized,gromov2019towards,you2019fractonic,radzihovsky2020fractons,devakul2018fractal,pai2019fracton,shirley2018foliated,oh2023aspects,delfino20232d,aasen2020topological,williamson2019spurious,watanabe,you2018symmetric,you2020symmetric,gorantla20232+,seiberg2020exotic,gorantla2022global,seiberg2021exotic}. The restricted mobility of quasiparticles can be naturally understood in terms of a set of the higher moment (or charge multipole moment) conservation laws\cite{pretko2017generalized,pretko2017generalized,gromov2019towards,nguyen2020fracton,radzihovsky2020fractons}, which often arise as a consequence of an emergent higher-rank gauge theory. For example, the simplest higher-rank gauge theories feature the conservation of both charge and dipole moment, which immobilizes individual charges but allows for the motion of stable dipolar bound states. Parallel to this development, similar approaches have been proposed for understanding fractonic phases using a \textit{symmetry-gauging principle} by considering certain exotic higher-form symmetries \cite{qi2021fracton,hirono2024symmetry,rayhaun2023higher}, subsystem symmetries \cite{you2018subsystem,devakul2018fractal,devakul2018strong}, or global symmetries that act on quasiparticles with position-dependent charge \cite{williamson2019fractonic,pace-wen,tantivasadakarn2021hybrid}.

In addition to shedding light on the field of fractons, the higher-rank gauge theory formalism has also drawn unexpected connections between fracton physics and other areas of physics, such as elasticity theory\cite{pretko2018symmetry,pretko2018fracton,gromov2024colloquium,kumar2018symmetry,nguyen2020fracton,radzihovsky2020fractons,radzihovsky2022lifshitz} and superfluid vortices\cite{doshi2020vortices,du2024noncommutative}. Ref.~\cite{pretko2018fracton} initiates the fracton-elastic duality to manifest that elasticity theory in two dimensions is dual to a symmetric tensor gauge theory, which explains the restricted mobility of disclinations and dislocation defects in solids. This type of duality has been extended to explore phase transitions in quantum solids such as supersolids~\cite{pretko2018symmetry,kumar2018symmetry} and vortex crystals~\cite{nguyen2020fracton}. In a similar vein, Ref.~\cite{doshi2020vortices} delineates that the vortex dynamics in superfluids exhibit fracton behavior, conserving both the total dipole moment and the trace of the quadrupole moment in terms of vorticity. This observation agrees with the fact that an isolated vortex is immobile, but vortex dipoles can move perpendicular to their dipole moment.

Among the novel and diverse aspects covered in the fracton literature, this review will present a pedagogical and chronological overview, with a particular focus on \textit{gapless fracton liquids}. We refer to these systems as `liquids' due to their compressibility, characterized by gapless excitations and algebraic decay of correlations. Their low-energy dynamics are primarily driven by collective excitations arising from emergent gauge fluctuations, mirroring the fundamental principles observed in Maxwell's electromagnetic theory. These are termed `fracton liquids' because their emergent charge and flux adhere to a higher-moment conservation law, thus lacking the mobility to move independently. In Sec.~\ref{sec:gauge}, we review the formulation of fracton liquids in 3D from emergent symmetric tensor gauge theory. The exploration of the theoretical framework of fracton liquids is followed by a discussion on their material realization in Yb-based breathing pyrochlore lattices (Sec.~\ref{sec:realize}) and close-packed tiling systems (Sec.~\ref{sec:close}). The review concludes by summarizing a general protocol (Sec.~\ref{sec:gen}) for creating and manipulating emergent tensor gauge theories within the realm of frustrated magnetism.

\section{Fracton liquid from tensor gauge theory}\label{sec:gauge}

In the primary exploration of fracton topological order, significant advances have emerged from exactly solvable lattice models. While the term \textit{fracton} was first proposed and introduced in Ref.\cite{Vijay2015-jj,Vijay2016-dr,Chamon2005-fc,Haah2011-ny,yoshida2013exotic,williamson2016fractal} within the context of stabilizer codes, discrete Lattice Gauge Theory, and duality, the origins of this thrust can be traced back to the exploration of symmetric U(1) `tensor gauge theory' in condensed matter systems\cite{xu2006gapless,xu2010emergent,rasmussen2016stable,pretko2017subdimensional,pretko2017generalized,pretko2017emergent,du2022volume,slagle2019symmetric,gorantla20232+,seiberg2020exotic,gorantla2022global,seiberg2021exotic} (also referred to as `higher-rank gauge theory'; we will use these two nomenclatures interchangeably without discrepancy throughout this review). The primary motivation was to pursue stable gapless algebraic liquids with power-law correlations emanating from symmetric U(1) tensor gauge fields, now denoted as \textit{fracton liquids}. They are treated as \textit{liquid} because the system is compressible with gapless excitations and algebraic decay of correlation, and intriguingly, the low-energy degrees of freedom are attributed to some collective excitations from emergent gauge fluctuations whose equations of motion resemble those of photons in a Maxwell theory. We embellish them as \textit{fracton liquids} because the emergent charge and flux are subjected to a higher-moment conservation law, thus lacking the mobility to move independently.

In Ref.~\cite{xu2006gapless,xu2010emergent,rasmussen2016stable}, the authors introduced a variety of lattice models that can be microscopically realized within interacting quantum spin Hamiltonians or hardcore boson systems. These systems feature symmetric tensor-like gauge theories, with charge/monopole excitations displaying restricted mobility due to higher-moment conservation laws.
In the pursuit of fracton liquids from concrete lattice models, the implementation of strong interactions plays a crucial role as it engenders the gauge symmetries. The underlying principle is that many-body interactions among hardcore bosons (or spins) enforce the total boson number within each local cluster to be fixed at low energies, thus generating a local symmetry of conserved charges. This local charge conservation can be interpreted as an emergent gauge symmetry (akin to thew Gauss's law) within the low-energy Hilbert space. Additional perturbations and quantum fluctuations can trigger resonances between different gauge symmetry sectors. These phenomena are dynamically interpreted as the quantum fluctuations of an emergent gauge field at the infrared (IR) scale, echoing the principles of generalized Maxwell theory.

The emergence of gauge theory in concrete lattice models can historically be traced back to the prototypical quantum spin ice systems, descended from rare-earth pyrochlore magnets like Dy$_2$Ti$_2$O$_7$ and Ho$_2$Ti$_2$O$_7$. This system is characterized by emergent electromagnetism and photon-like excitations at low energy, arising from local spin number conservation and quantum fluctuations\cite{castelnovo2008magnetic,hermele2004pyrochlore,ross2011quantum}. In 3D spin ice materials composed of regularly corner-linked tetrahedra of magnetic ions, the interactions are primarily governed by the classical antiferromagnetic Ising coupling \(S^z_iS^z_j\) between pairs of sites in each corner-linked tetrahedron. This interaction imposes an intriguing rule on the four spins in each tetrahedron, enforcing an all-in-all-out configuration so that the total number of \(S_z\) is zero for all tetrahedra. Such rule creates an extensive array of configurations within the ground state manifold, as the all-in-all-out configuration can be treated as a divergence-free `Gauss's law' for the electric line. Consequently, the spin configurations in the ground state resemble closed electric loops within the charge vacuum background. By considering quantum fluctuations, such as the \(S^+_iS^-_j\) interaction that flips the \(S^z\) spin, different closed-loop patterns can resonate with each other. This resonance enables fluctuations between distinct electric line configurations, effectively mirroring magnetic field fluctuations. The low-energy spin excitation bears resemblance to the electromagnetic fluctuation observed in 3D Maxwell's theory, featuring a gapless spectrum similar to photon excitations\cite{hermele2004pyrochlore}.

\begin{figure}[h!]
\includegraphics[width=4.8in]{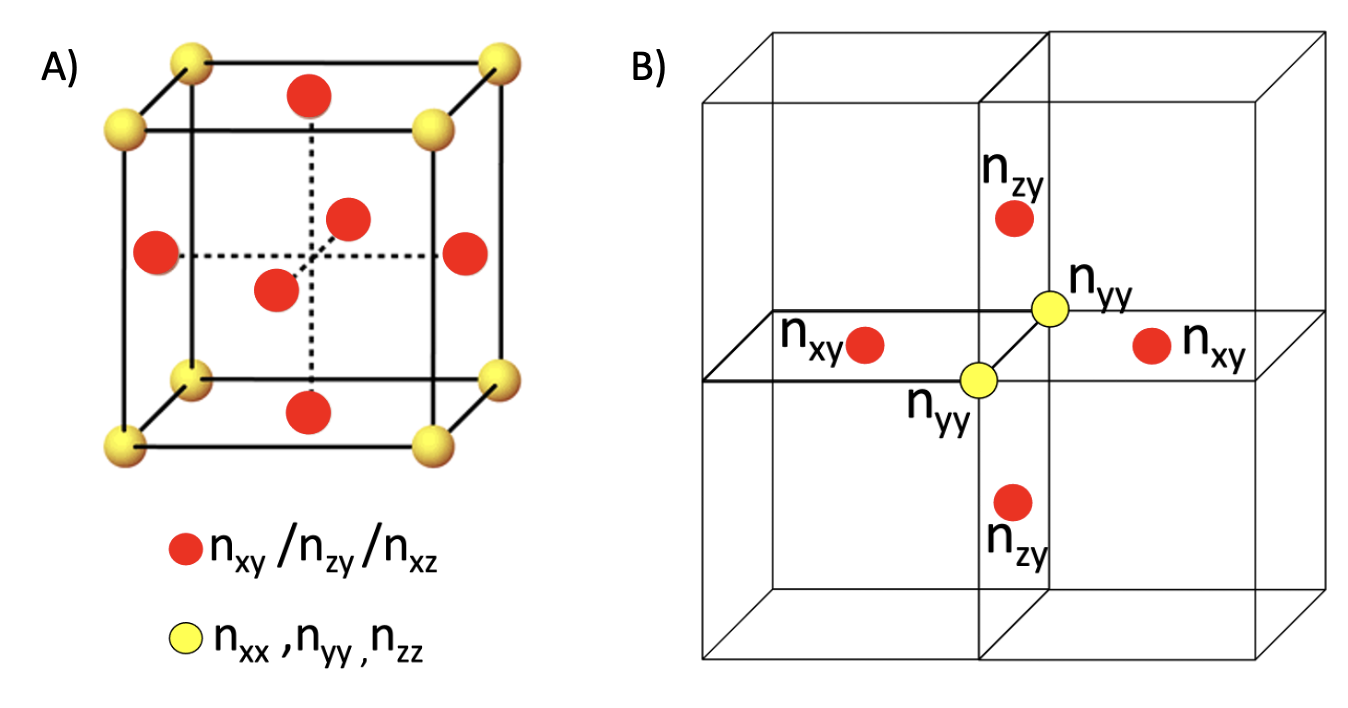}
\caption{A) The distribution of orbital boson degrees of freedom on an fcc lattice: There are three orbital bosons, \(n_{xx}\), \(n_{yy}\), and \(n_{zz}\), at each vertex (yellow dot). The orbital boson \(n_{ij}\) resides at the center of the plaquette (red dot) on the i-j plane.
B) An illustration of one of the vector conservation laws, \(\partial_i n_{iy} = 0\), which fixes the number of orbital bosons in this cluster (adjacent to the center y-link).} \label{3dlattice}
\end{figure}

\subsection{Rank-2 fracton liquid from interacting bosons on the FCC lattice}

Leveraging the principles of quantum spin ice,  Ref.\cite{xu2006gapless,xu2010emergent,rasmussen2016stable} propose an innovative construction of fracton liquid from interacting boson models on an fcc lattice, whose IR theory manifests as a \textit{rank-2 symmetric tensor gauge theory} featuring soft graviton excitations. The lattice model setp up is depicted in Fig.~\ref{3dlattice}: The vertices of the fcc lattice accommodate three orbitals of bosons, labeled \(n_{xx}\), \(n_{yy}\), \(n_{zz}\), while the center of each plaquette on the i-j plane hosts another flavor of boson, labeled \(n_{ij}\).

The microscopic Hamiltonian begins with a constraint that effectively imposes the conservation of the total number of bosons in each cluster illustrated as Fig.~\ref{3dlattice} (Here, we bipartition the lattice and modify the boson number with a sublattice factor \(n_{ij}(r) \rightarrow (-1)^{x+y+z}n_{ij}(r)\), such that the boson charge on one of the sublattices is treated as a hole). 
This conservation law can be compactly expressed in the continuum as follows:
\beqn
\partial_{x}n_{xx} + \partial_y n_{xy} +
\partial_{z}n_{xz} = 0,  \partial_{x}n_{xy} + \partial_y n_{yy} +
\partial_{z}n_{yz} = 0,  \partial_{x}n_{xz} + \partial_y n_{yz} +
\partial_{z}n_{zz} = 0.\label{cons3}\eeqn 

Although the microscopic realization for engendering such interactions, which enforce a fixed number of bosons per cluster, was unclear at the time of its proposal (the material realization of this model will be discussed in Sec.~\ref{sec:realize}), the essential idea is that the low-energy subspace is governed by a local constraint. This constraint is reminiscent of a generalized Gauss's law, endowing the low-energy space with an emergent gauge symmetry.

The many-body Hilbert space is delineated by the occupancy numbers of these bosons on the vertices and plaquettes, labeled as \(n_{ij}\) (spanning \(i, j\) across \(x, y, z\), with \(n_{ij} = n_{ji}\)), and their canonical conjugate phases \(\theta_{ij}\). Inspired by insights from quantum spin ice, one can interpret the boson number operator for each orbital as a symmetric rank-2 tensor electric field, denoted as \(E_{ij} \sim n_{ij}\). Similarly, the phase \(\theta_{ij}\) assumes the role of the gauge potential, with \(A_{ij} \sim \theta_{ij}\). It is noteworthy that the electric field \(E_{ij}\) assumes integer values, and \(A_{ij}\) adheres to a compact modulo \(2\pi\). In this formalism, the fixed number of bosons in each cluster, defined in Eq.~\ref{cons3}, can be mapped into a vector conservation law from a generalized electromagnetic perspective:
\beqn \label{lc2}
\partial_i E_{ij} = \rho_j, \eeqn 
In the context of the original lattice model, characterized by a fixed number of bosons per cluster as defined in Eq.~\ref{cons3}, this scenario corresponds to the charge vacuum condition with \(\vec{\rho}=0\).
This vector charge theory obey two types of constraints:
\begin{equation}
\int\vec{\rho}\, = \textrm{constant}\,\,\,\,\,\,\,\,\,\,\,\,\,\,\int\vec{x}\times\vec{\rho}\, = \textrm{constant}
\label{cons}
\end{equation}
reflecting the conservation of both the vector charge and its angular charge moment. To align with these conservation laws\cite{pretko2017subdimensional,pretko2017generalized,pretko2017emergent}, the vector charges behave like sub-dimensional particles (denoted as fractons) with restricted mobility, capable of hopping exclusively in the direction parallel to their charge vector. 

This vector conservation law also defines the gauge transformation of the gauge potential:
\beqn \label{lc3}
A_{ij} \rightarrow A_{ij} +
\frac{1}{2}\left(\partial_i \lambda_j + \partial_j
\lambda_i\right) \eeqn 
Intriguingly, this gauge transformation mirrors aspects of linearized gravity when the geometry metric tensor \( \eta_{ij} \) is substituted with \( A_{ij} \)\cite{xu2010emergent,pretko2017emergent}.

When conservation of bosons in each cluster is imposed as a gauge symmetry in the low-energy Hilbert space, as outlined in Eq.~\ref{lc2}, the quantum fluctuations that survives at infrared (IR) scales are required to be gauge-invariant. Through straightforward algebra, we can determine the lowest-order gauge-invariant quantity, denoted as the tensor magnetic flux \( B_{ij} \), which takes the form:
\beqn  B_{zz} =
\epsilon_{zab}\epsilon_{zcd}\partial_a\partial_cA_{bd}, ~~B_{yy}
=\epsilon_{yab}\epsilon_{ycd}\partial_a\partial_cA_{bd} ,
B_{xx} =
\epsilon_{xab}\epsilon_{xcd}\partial_a\partial_cA_{bd},\cr ~~B_{yz}
= \epsilon_{yab}\epsilon_{zcd}\partial_a\partial_cA_{bd},
B_{xz} =
\epsilon_{xab}\epsilon_{zcd}\partial_a\partial_cA_{bd}, ~~B_{xy}
= \epsilon_{xab}\epsilon_{ycd}\partial_a\partial_cA_{bd}. 
\label{doublecurl}\eeqn 
This magnetic field is a rank-2 symmetric tensor, defined by the double curl of \( A_{ij} \).
The compact form of the magnetic flux \( B_{ij} \) is:
\beqn \label{b2} B_{ij} =
\epsilon_{iab}\epsilon_{jcd}\partial_a\partial_c A_{bd}\eeqn

Here are a few remarks on the physical implications of such a tensor magnetic flux operator. The magnetic flux in this theory is a line-like object that transforms as a tensor.
Beyond the strong interactions within each cluster that impose a constraint on the total boson number, one can consider the quantum fluctuations among different orbital bosons. These fluctuations facilitate their hopping and exchange between sites, as indicated by
\beqn  H_t=-t_{r,r'}\cos(\theta_{ab}(r)-\theta_{cd}(r'))+... \eeqn
However, the motion of a single boson through hopping and exchange in space is significantly suppressed due to the emergent gauge symmetry with fixed number of bosons in each cluster. Nevertheless, considering the effects of higher-order perturbations, one can demonstrate that the spatial fluctuation of a bound state of multiple bosons, in a manner that preserves the cluster's total boson number, is still allowed. Following some straightforward algebra, one can ascertain that the leading-order perturbation term can be expressed as \(\cos(B_{ij})\), achieved by reinterpreting the compact phase \(\theta_{ab}\) in terms of the tensor gauge potential \(A_{ab}\). This perspective is also supported by gauge symmetry considerations. As the low-energy Hilbert space manifests an emergent gauge symmetry, the quantum fluctuation should originate from gauge-invariant quantities, which in this case would be the magnetic flux operator.

\subsection{Maxwell theory and instanton events}

We now formulate a microscopic Hamiltonian for this theory, drawing inspiration from Maxwell's theory of electromagnetism:
\begin{align}\label{maxtensor}
 H = U\sum_r E_{ij}^2 + K\sum_r \cos(B_{ij}) \approx U\sum_r E_{ij}^2 + K B_{ij}^2  
\end{align}
This expression encapsulates the self-energy density of the rank-2 electromagnetic field, which was first introduced in Ref.~\cite{rasmussen2016stable,pretko2017generalized} dubbed as \textit{Rank-2 gauge theory}.
The Hamiltonian reveals a gapless `soft graviton' mode characterized by a quadratic dispersion, \(\omega \sim k^2\). In the last step, we employed a spin-wave expansion of the compact magnetic flux operator as an approximation. However, given the compact nature of \(A_{ij}\), an instanton event can occur, resulting in a shift of \(B_{ij} \rightarrow B_{ij} + 2\pi\). Such instanton tunneling events have the potential to destabilize the gapless mode. This phenomenon mirrors what is observed in compact U(1) gauge theory in \(2+1\) dimensions, where flux instantons induce charge confinement.
To explore the impact of instanton tunneling\cite{rasmussen2016stable}, one can depict the electric field using a discrete height field \(h_{ij}\). This is achieved by expressing \(E_{ij}\) as the 2-curl of the height field \(h\):
 \beqn E_{ij} =
\epsilon_{iab}\epsilon_{jcd}\partial_a\partial_c h_{bd}. \eeqn 
The vector conservation law specified in Eq.~\ref{lc2} naturally arises for smooth configurations of the \(h_{ij}\) field. Importantly, \(h_{ij}\) undergoes the same gauge transformation as the gauge potential \(A_{ij}\), and serves as the canonical conjugate partner to the magnetic field tensor \(B_{ij}\). The operator \(e^{i 2\pi h_{ij}}\) triggers an instanton tunneling event, functioning as the vertex operator that proliferates the magnetic field tensor by a multiple of \(2\pi\). However, due to the special gauge invariance associated with \(h_{ij}\), such a vertex operator is prohibited at low energy, thus rendering the absence of instanton events at the infrared Gaussian fixed point, as it breaks gauge symmetry. The irrelevance of the instanton event in this context guarantees the stability of the quadratic gapless mode \(\omega = k^2\), culminating in the stability of a deconfined U(1) tensor gauge theory.

A brief comment on the duality structure of this gauge theory. From the magnetic field tensor defined in Eq.~\ref{doublecurl}, it is evident that the theory establishes a vector conservation law for the magnetic monopole:
\begin{equation} \label{lcm}
\partial_i B_{ij} = 0, \end{equation}
which is reminiscent of the vector charge conservation law in Eq.~\ref{lc2}. Given the similarities in the gauge structures of charge and monopole, the deconfined gauge theory exhibits self-duality, with identical gauge symmetries on both sides of the duality. This is further evidenced by the fact that the height field \(h_{ij}\), which generates instanton events for the magnetic flux, undergoes the same gauge transformation as the gauge potential \(A_{ij}\). The self-dual structure has significant implications for the stability of this gauge theory, as it imposes constraints on instanton events from a symmetry perspective\cite{pretko2017subdimensional}. Once again, due to the compactness of \(A\), we need to consider vector compact monopole patterns \(\partial_i B_{ij} = 2\pi n^j\). Employing the same arguments used for vector charge, these vector monopoles $\vec{n}$ are subjected to the conservation laws of both vector monopole and angular moment of the monopole. Such conservation of higher moments in monopole restricts the proliferation of magnetic instantons, thereby ensuring that the theory is deconfined.

Finally, we include a brief note to clarify that the tensor gauge theory introduced here does not belong to the category of \textit{higher-form} gauge theory, which specifically refers to antisymmetric tensor objects that couple to loop or membrane currents. The symmetric tensor gauge theories in Eq.~\ref{maxtensor} (also dubbed as higher-rank gauge theory in some literature)
present a novel class of stable fracton liquids originating from higher moment conservation laws. The gapless modes are triggered by gauge fluctuations and are protected against any microscopic perturbation to the Hamiltonian, regardless of symmetry considerations.

\subsection{Properties of the fracton liquid from tensor gauge theory}

So far, we have reviewed a novel type of fracton liquid emerging from tensor gauge theories. As fracton liquids, their salient features lie in the fact that the theory is gapless without symmetry breaking\footnote{Here, symmetry breaking denotes spontaneous breaking of global symmetries. We do not consider more exotic cases like higher-form symmetry breaking.} and stable against perturbations. The gaplessness is protected by emergent gauge invariance, enforced by a local constraint on the low-energy Hilbert space. Thus, it might be meaningful to look into the correlation function of the electric tensor and see how it decays in spacetime.
Another aspect of paramount importance is the experimental signatures that differentiate fracton liquids. As the fracton liquid introduced here can be effectively described as a symmetric tensor gauge theory with a higher moment conservation law, a pertinent question arises: How can we ascertain that these theories are distinct from previously understood 3D vector gauge theories, such as U(1) spin liquids in quantum spin ice? Is there an experimental `smoking gun' that fundamentally distinguishes them?

Ref.~\cite{rasmussen2016stable} benchmarks the unique feature of this symmetry tensor gauge theory, characterized by a power-law correlation in \(E_{ij}\) (which originates from the orbital boson charge density operator), which diminishes as \(\sim \frac{1}{r^5}\) over long distances\cite{xu2006gapless}, featuring an exponent greater than that observed in the quantum spin ice phase. Likewise, Pretko\cite{pretko2017subdimensional} considers electrostatic confinement by examining the energy stored in the electric field under a static vector charge background and finds that the electric tensor \(E_{ij}\) scales as \(1/r^2\), similar to a normal Coulomb field.
 
To visualize the gauge structure of the fracton liquid, one can further calculate the electric correlation in momentum space, which exhibits momentum-dependent singularities at low temperatures, denoted as pinch points\cite{prem2018pinch}:
\begin{align}\label{pinch}
\langle E^{ij}(q)E^{k\ell}(-q) \rangle\propto  \bigg[\frac{1}{2}(\delta^{ik}\delta^{j\ell} + \delta^{i\ell}\delta^{jk})  + \frac{q^iq^jq^kq^\ell}{q^4}  - \frac{1}{2}\bigg(\delta^{ik}\frac{q^jq^\ell}{q^2} + \delta^{jk}\frac{q^iq^\ell}{q^2} + \delta^{i\ell}\frac{q^jq^k}{q^2} + \delta^{j\ell}\frac{q^iq^k}{q^2}\bigg)\bigg]. 
\end{align}
The emergence of pinch-point singularities in the correlation function is attributed to the second term, a result of the projection into the gauge sector. As one approaches the origin, the ratio within this term exhibits varying limits depending on the angular direction, contributing to the distinctiveness of the singularities. Moreover, this correlation function displays a characteristic four-fold symmetry, in stark contrast to the two-fold symmetry observed in 3D quantum spin ice. Particularly noteworthy is that the pinch point singularity at finite momentum can be detected through neutron scattering data.

The dynamics of the fracton liquid are encapsulated by a set of generalized Maxwell's equations\cite{rasmussen2016stable}:
\beqn
&\partial_i E_{ij} = 0, ~ ~\partial_i B_{ij} = 0,~
~\cr\cr
&\partial_t E_{ij} -
\kappa\epsilon_{iab}\epsilon_{jcd}
\partial_a\partial_c B_{bd} = 0,~~
\partial_t B_{ij} + \kappa\epsilon_{iab}\epsilon_{jcd}
\partial_a\partial_cE_{bd} = 0.\label{maxwell}\eeqn 
The first two equations establish Gauss's law for tensor electric and magnetic fields. The subsequent two explain the generation of tensor electric (magnetic) fields through time-varying magnetic (electric) fields. This process leads to the formation and propagation of a rank-2 electromagnetic wave across space.

\subsection{Another variety of fracton liquids: Traceless symmetric tensor gauge theory}\label{sec:traceless}

The inherent tensor nature of higher-rank gauge theories allows us to incorporate additional local constraints with additional gauge symmetry, paving the way for more exotic fracton liquids. For instance, starting with the previously mentioned rank-2 gauge theory, we introduce a traceless condition for the electric tensor:
\beqn \label{lc34} E = \sum_i E_{ii} = 0
\eeqn This traceless constraint generates another gauge transformation \beqn A_{ij}
\rightarrow A_{ij} + \delta_{ij}\lambda \eeqn 
With the introduction of this additional constraint, the leading order gauge-invariant operator is identified and treated as the magnetic field operator: \beqn \label{q2} Q_{ij} = \epsilon_{ikl}\left(B_{jl} -
\frac{1}{2}\delta_{jl}B\right), \eeqn  
The Maxwell term of the tensor gauge theory reads: \beqn H = U\sum_r
E_{ij}^2 + K\sum_r Q_{ij}^2. \label{z3}\eeqn 
The theory exhibits a self-dual structure in which the electric and magnetic tensors are dual to each other, sharing a similar gauge structure. Originally proposed in Ref.~\cite{rasmussen2016stable}, this theory features a stable gapless liquid phase where the low-energy excitations display cubic dispersion, \(\omega \sim k^3\).

\subsection{Generalized higher-rank gauge theory}

The process of creating fracton liquids originating from emergent tensor gauge theories can be effectively extended to various higher-rank gauge fields through the application of the `fracton-gauge principle' outlined in Ref.~\cite{pretko2017generalized,rasmussen2016stable,pretko2018fracton,gromov2019towards}. This methodology involves several crucial steps:
i) Identifying a generalized symmetry that conserves the charge multipole moment (such as dipole, quadrupole, octupole moments, or a polynomial combination of them).
ii) Gauging the symmetry by coupling it with a local higher-rank gauge potential.
iii) Defining a generalized Gauss's law that reflects the specific charge multipole conservation.
iv) Defining the magnetic field operator by identifying the lowest-order gauge-invariant operator, thereby establishing a Maxwell-type description of the low-energy dynamics.
v) Given that the emergent gauge theory is compact, we need to examine the relevance of instanton events that potentially lead to confinement. Notably, it is sometimes necessary to consider higher-order instanton operators, such as instanton dipoles and quadrupoles with discontinuity field patterns, as they may turn out to be more relevant due to the consequences of UV-IR mixing.
In Table 1, we include a summarization of fracton liquids from generalized higher-rank gauge theory in 3D, originally explored in Ref.\cite{rasmussen2016stable} and enriched from a fracton perspective in Ref.\cite{pretko2017generalized}.
\begin{center}
\begin{table}[h!]
\begin{tabular}{| l | l | l | l|}
\hline
Rank of theory & Local constraint & Gapless dispersion & Heat capacity  \\ \hline
\multirow{1}{*}{$n=1$} & $\partial_i E_i = 0$ & $\omega \sim k$& $C_v \sim T^3$  \\ \hline
\multirow{3}{*}{$n=2$} & $\partial_i E_{ij} = 0$ & $\omega \sim k^2$ & $C_v \sim T^{\frac{3}{2}}$  \\
& $\partial_i \partial_j E_{ij} = 0$ & $\omega \sim k$ & $C_v \sim T^3$  \\
& $\partial_{ij} E_i = 0,E_{ii} = 0$ & $\omega \sim k^3$ & $C_v \sim T$ \\ \hline
\multirow{2}{*}{$n=3$} & $\partial_i E_{ijk} = 0$ & $\omega \sim k^3$ & $C_v \sim T$  \\
& $\delta_{ij} E_{ijk} = 0, \partial_i E_{ijk} = 0$ & $\omega \sim k^5$  & $C_v \sim T^{\frac{3}{5}}$ \\ \hline
\end{tabular}
\caption{Summary of higher-rank gauge theories in 3D.}
\end{table}
\end{center}

Building on this foundation, Ref.\cite{gromov2019towards,gromov2020fracton,bulmash2018generalized} proposed the fracton gauge principle alongside the charge multipole algebra approach, offering a framework to articulate a broader class of higher-rank gauge theories. This approach involves starting with ungauged quantum matter fields characterized by exotic conservation laws of charge multipole moments and gauging the theory by coupling them with a local gauge potential. Furthermore, Ref.~\cite{ma2018fracton,bulmash2018higgs,delfino2023anyon} elucidates that when the symmetric U(1) tensor gauge theory undergoes Higgsing, combined with the condensation of partial confinement mechanisms, this process transforms the gapless fracton liquid into a fully gapped fracton topological ordered phase known as the fracton stabilizer code whose ground state degeneracy scales sub-extensively with system size.

From an alternative perspective, Ref.\cite{hirono2024symmetry} demonstrated that various gapless fracton liquids, echoing the properties of the tensor gauge theories mentioned earlier, can result from the spontaneous symmetry breaking (SSB) of nonuniform higher-form symmetries\cite{qi2021fracton,rayhaun2023higher}, where the associated conserved charges permute nontrivially under spatial translations. 
Under this framework, the worldlines of particles become immobile, as they are considered charged objects subject to 1-form symmetries. Through this approach, instead of identifying a special form of charge multipole symmetry and delineating its corresponding gauged theory, it allows us to identify exotic fracton liquids whose quasiparticle mobility restrictions are precisely defined by the chosen commutation relations between charges and translations\cite{williamson19,pace-wen}.
In a parallel thread, attention has been directed towards the twisted tensor gauge theory, where the general Maxwell theory for higher-rank electromagnetic fields is enhanced by an additional Chern-Simons type coupling\cite{you2020fractonic,you2019multipolar,ma2022fractonic,huang2023chern,prem2017emergent,shirley2020fractonic,tantivasadakarn2021hybrid,delfino20232d}. With this modification, the flux excitation carries a gauge charge and thus exhibits nontrivial sub-dimensional braiding statistics in 3D\cite{you2020symmetric,song2024fracton,sullivan2021fractonic}.

\section{Yb-based breathing pyrochlores: A window to fracton liquids} \label{sec:realize}

The wide variety of proposals for fracton liquids, originating from microscopic lattice models with emergent tensor gauge theories, underscores the necessity for physically realistic materials capable of hosting such states\cite{pretko2017fracton,Slagle2017-la,you2018majorana,you2019emergent,hsieh2017fractons,yan2019rank,benton2021topological,han2022realization,han2022non}. Discovering novel higher moment conservation laws of nature arising from frustrated magnetism is not only of great intellectual interest but also holds significant potential for material applications.
Motivated by the rapid development of Kitaev materials discovered in systems with correlated spin-orbit coupling, it has been proposed in Ref.~\cite{yan2019rank} that Yb-based iridate materials, possessing a quantum spin ice-like structure, could potentially realize stable fracton liquids characterized by rank-2 symmetric traceless gauge theories reviewed in Sec.~\ref{sec:traceless}. In this section, we will introduce and elaborate on the material proposal in Ref.~\cite{yan2019rank}, which suggests a variety of higher-rank spin liquids derived from frustrated magnetic materials.

\begin{figure}[h!]
    \centering
   \includegraphics[width=0.45\textwidth]{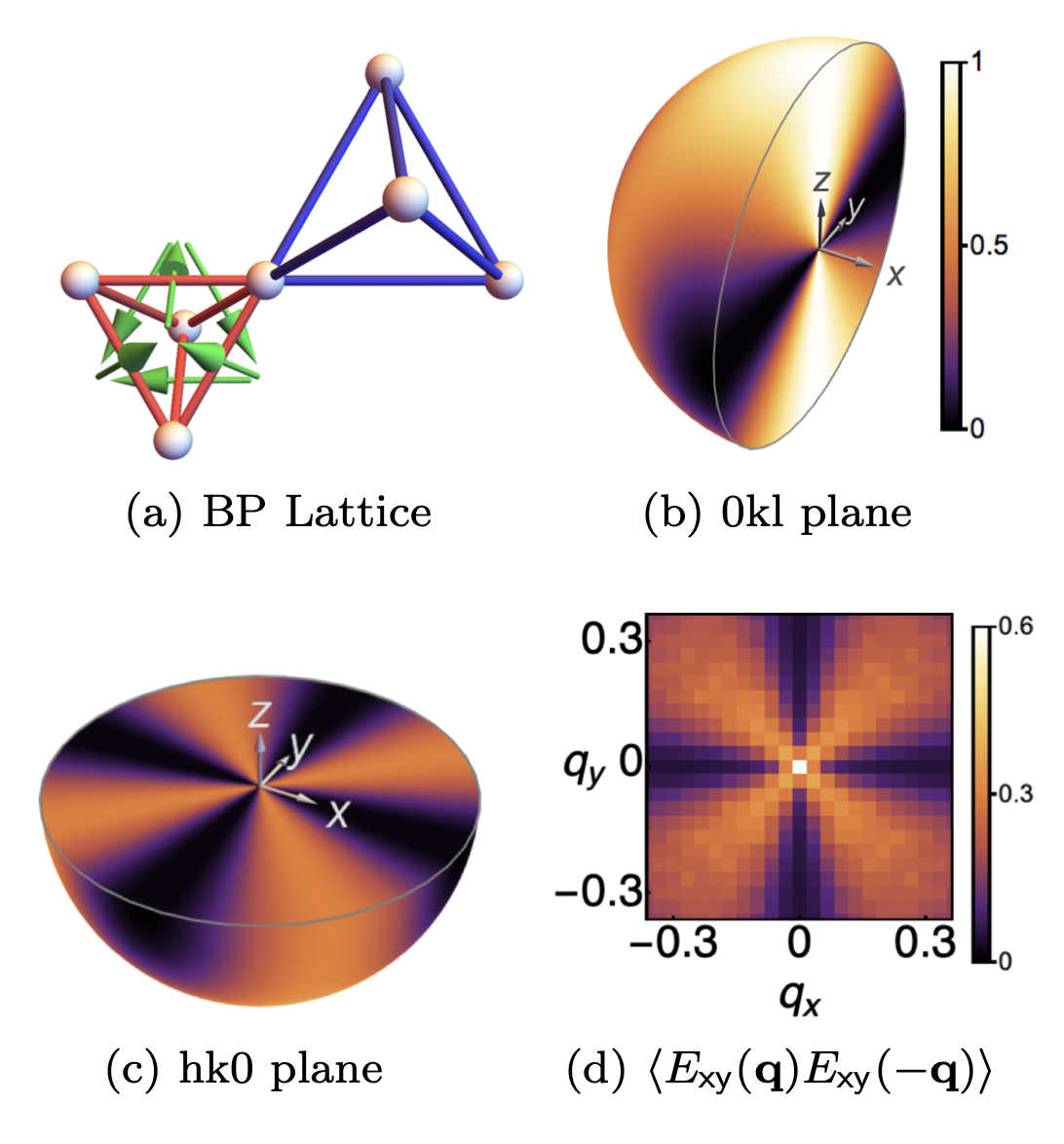}
    \caption{ a) Breathing pyrochlore lattice, with A- and B-sublattice tetrahedra of different sizes. The vectors associated with Dzyaloshinskii-Moriya interactions on the A-sublattice are illustrated with green arrows. 
b) Prediction of R2–U1 theory from the electric correlation function \(\langle E_{xy}(q) E_{xy}(-q)\rangle\), showing a 2-fold pinch point. 
c) A 4-fold pinch point in the perpendicular plane. 
d) Classical Monte Carlo simulation of the breathing pyrochlore model for the correlation. The figure is reproduced from Ref.~\cite{yan2019rank}}
    \label{fig:R2SL}
\end{figure}

Frustrated magnetism provides a fertile ground for exploring novel quantum states. The interplay between geometric frustrations and quantum fluctuations leads to highly-entangled spin patterns, fostering long-range correlations and hidden orders. Across a broad spectrum of magnetic materials, frustration can give rise to extensive degenerate ground state configurations with conserved spin numbers in each local cluster, interpreted as the emergence of gauge symmetries.
A prime example of this phenomenon is the quantum spin ice, reminiscent of the \(U(1)\) gauge theory on the pyrochlore lattice\cite{castelnovo2008magnetic,hermele2004pyrochlore,ross2011quantum}. Its low-energy excitations remarkably mimic elements of Maxwell electrodynamics, including photons, electric charges, and magnetic monopoles.

In Ref.~\cite{yan2019rank}, it was outlined that the traceless symmetric rank-2 \(U(1)\) fracton liquid, as detailed in Sec~\ref{sec:traceless}, naturally emerges in Yb-based breathing pyrochlores. The material is primarily governed by a Heisenberg antiferromagnet (HAF) on a breathing-pyrochlore lattice and influenced by weak Dzyaloshinskii-Moriya (DM) interactions, as illustrated in Fig.~\ref{fig:R2SL}. 
The effective Hamiltonian for the system is captured by the breathing Heisenberg antiferromagnetic interactions on both the A- and B-tetrahedra, and negative Dzyaloshinskii-Moriya (DM) interactions only on the A-tetrahedra:
\begin{equation}\label{EQN_S_hamiltonian}
\mathcal{H} =
\sum_{\langle ij \rangle\in \text{A}} \left[J_A S_i \cdot S_j 
+
D_A\hat{d}_{ij}\cdot( S_i \times S_j )\right ] 
+ 
\sum_{\langle ij \rangle\in \text{B}} \left[J_B S_i \cdot S_j 
)\right ].
\end{equation}
where $\langle ij \rangle \in \text{A(B)}$ denotes
nearest neighbour bonds belonging to the A(B)-tetrahedra.
Vectors $\hat{\bf d}_{ij}$ are bond dependent whose definition is detailed in Ref.~\cite{yan2019rank}.
The Heisenberg parameters \(J_{A/B}\) and the DM interactions \(D_{A}\) generate various frustrations and thus trigger a rich phase diagram with competing orders. By introducing coarse-grained fields that transform as irreducible representations of the lattice symmetry, Ref.\cite{yan2019rank} demonstrates that in the following parameter regime,
\begin{eqnarray}
J_A\ ,\ J_B>0\ ,\ D_A<0 \; . 
\end{eqnarray}
The effective theory of the coarse-grained fields exhibits an emergent gauge symmetry that can be characterized by the following traceless vector conservation law:
\beqn 
\partial_i E_{ij} = 0, ~ \sum_iE_{ii} = 0,\eeqn 
Which renders the rank-2 symmetric traceless tensor gauge theory reviewed in Sec~\ref{sec:traceless}. At finite temperatures, such tensor gauge theory is stable within a finite parameter region, denoted as the [\mbox{R2--U1}] spin liquid phase in the spin liquid literature. To identify the [\mbox{R2--U1}] spin liquid and distinguish it from traditional quantum spin ice, numerical simulations of the spin structure factor were conducted. Such spin structure factor, akin to the electric tensor correlator in Eq.~\ref{pinch}\cite{prem2018pinch}, is expected to display momentum-dependent singularities with four-fold symmetries.
\begin{figure}[h!]
    \centering
   \includegraphics[width=0.6\textwidth]{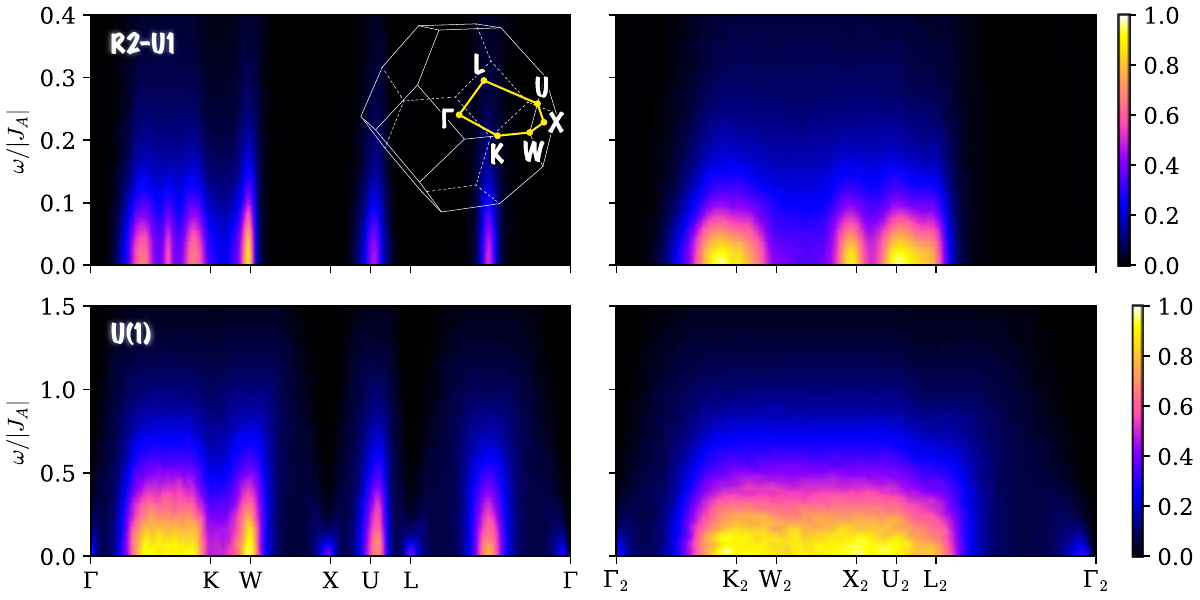}
    \caption{Energy dependence of dynamical structure factor along the $[hh0]$ momentum cut. The total dynamic structure factor (left column) and dynamic structure factor in the spin-flip (SF) channel (right column) are shown for the R2-U1 regime(top), compared to the spin ice regime(bottom).The figure is reproduced from Ref.~\cite{zhang2022dynamical}}
    \label{fig:kim}
\end{figure}
As shown in Fig.~\ref{fig:R2SL}, the presence of four-fold pinch point singularities in the equal-time spin correlations —highlighted as a smoking gun for this [\mbox{R2--U1}] spin liquid phase — can be precisely detected in polarized neutron scattering \cite{prem2018pinch}. In contrast, standard spin ice \cite{fennell2009experimental, benton2012seeing}, recognized as an emergent \(U(1)\) vector gauge theory, only shows two-fold pinch point singularities. Such a distinction can be traced back to the emergent gauge structure and conservation laws. The conventional quantum spin ice adheres to standard charge conservation in ordinary U(1) spin liquids, whereas the [\mbox{R2--U1}] spin liquid, embodying a rank-2 gauge theory, exhibits vector charge conservation that is traceless and symmetric.

In a follow-up study\cite{zhang2022dynamical}, Kim et al. further investigate this R2-U1 spin liquid within the inelastic spin structure factor using classical finite-temperature Monte Carlo techniques and molecular dynamics. The result reveals the persistent presence of the four-fold pinch point singularities in the dynamic structure factor at finite temperature and convinced that the dynamical signatures distinguishing the [\mbox{R2--U1}] spin liquid from quantum spin ice are highly distinctive, as depicted in Fig.~\ref{fig:kim}. These findings provide essential insights for upcoming inelastic neutron scattering experiments on magnetic materials, such as Yb-based breathing pyrochlores, and set a new benchmark for the study of different variations of higher-rank spin liquids.

\section{Fracton liquids from close-packed tiling problems}\label{sec:close}

The close-packed tiling problem has been thoroughly explored due to its broad applications, from spin glass and jamming to macromolecular contexts. Its low-energy degrees of freedom are generated by the local constraint of close-packed configurations, with dimer, trimer, and plaquette arrangements tiled around each site. At zero temperature, the system exhibits extensive entropy due to the large number of possible tiling patterns, and its equilibrium physics is dominated by entropy fluctuations. Ref.~\cite{xu2008resonating,pankov2007resonating} proposed the close-packed plaquette model on the cubic lattice, which can be viewed as a U(1) version of the X-cube theory where charge and flux are conserved on the submanifolds of 2D planes. Motivated by this approach, Ref.~\cite{you2022plaq} explores a broader class of close-packed tiling problems in 3D as a fertile ground to build 3D fracton liquids.

\begin{figure}[h!]
  \centering   \includegraphics[width=0.8\textwidth]{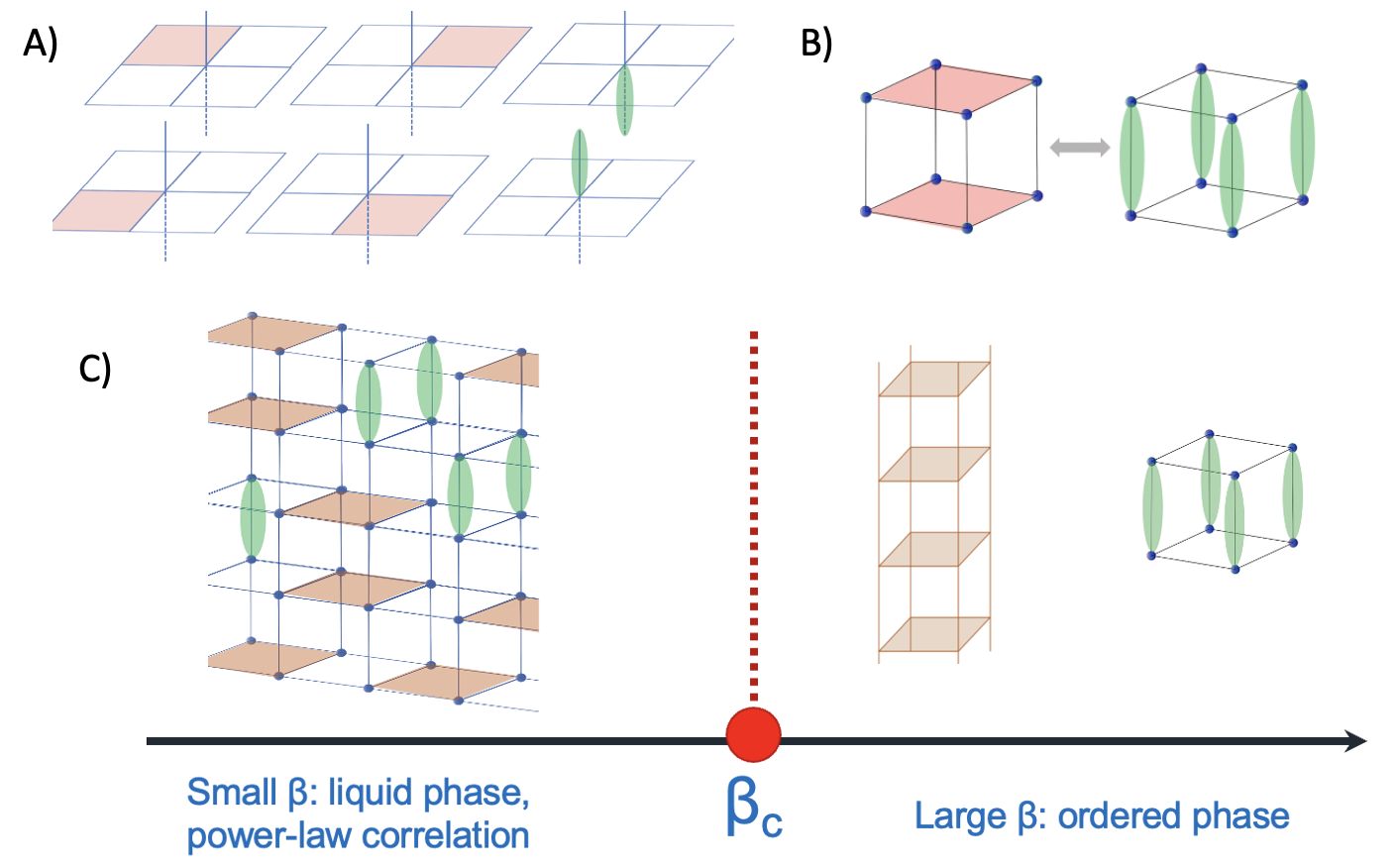}
  \caption{A) All possible close-packed patterns on a cubic lattice at each vertex. B) Thermal flippable patterns. C) Phase diagram: When \(\beta < \pi/2\), entropy fluctuations lead to a fracton liquid phase characterized by power-law correlations and quasi-long-range order. The transition from the liquid to the ordered phase is characterized by a fractonic Kosterlitz-Thouless transition.} 
\label{phasedia}
\end{figure}

The primary setup begins with a classical close-packed dimer-plaquette tiling problem on a cubic lattice, whose Hilbert space is spanned by distinct close-packed patterns: each site in the cubic lattice connects to either a dimer on the \(z\)-link or a plaquette on the \(xy\) plane as Fig.~\ref{phasedia}. This configuration of fully packed tiling patterns displays a macroscopic number of degeneracy with extensive entropy. Intriguingly, the close-packed dimer-plaquette patterns can be described through the lens of `higher-rank electrostatics', offering a unique perspective on their underlying gauge structure and dynamics\cite{shirley2018fractional,shirley2018foliated}.

To impose the close-packed constraint, one can define the dimer and plaquette occupancy on each link and plaquette as a higher-rank electric field\cite{you2019emergent,gromov2019towards},
\begin{align} 
E_{xy}=\eta P_{xy}, ~E_{z}=\eta D_{z}\ . \label{def}
\end{align}
$P_{xy}$ and $D_{z}$ refer to the number of plaquettes and dimers living on the x-y plaquettes and z-links, respectively. $\eta$ is the bipartite lattice factor. Based on this notation, the dimer-plaquette constraint can be interpreted as a generalized Gauss law,
\begin{align} 
\Delta_x \Delta_y E_{xy}+\Delta_z E_{z}=\eta (1-Q)
\label{cons}
\end{align}
Here, $\Delta_i$ represents the lattice difference. $Q$ denotes the number of monomers on the site, which should be taken to be zero for close-packed conditions. Intriguingly, monomer conservation occurs on all x-z and y-z planes, which means their dynamics are constrained exclusively to the z-direction.

In the context of the close-packing problem, the partition function is defined by summing over all possible configurations of plaquettes and dimers, each attributed equal Boltzmann weights. Due to the lack of energetic terms in the partition function, the system's free energy is derived entirely from entropy. When the \(E_{xy}\) and \(E_z\) fields are coarse-grained, flippable configurations (with coarse-grained average \(\bar{E}_{xy}, \bar{E}_z = 0\)), contribute to a larger ensemble of microscopic states and thus possess larger coarse-grained entropy compared to non-flippable configurations where \(\bar{E}_{xy}, \bar{E}_z \neq 0\).

To illustrate, let's consider a local flip on a cube that oscillates between two x-y plaquettes and four z-dimers, as depicted in Fig.~\ref{phasedia}. These flippable patterns demonstrate a zero average for both \(E_{xy}\) and \(E_z\), highlighting that the coarse-grained entropy effectively favors such configurations. The partition function of the system can be represented by a Gaussian theory of the electric field fluctuation as \(e^{-\beta((E_{xy})^2 + (E_z)^2)}\). This formulation quantifies the entropy associated with each coarse-grained field, with \(\beta\) acting as the stiffness parameter. When \(\beta\) is high, entropic fluctuations tend to promote strong ordering. Macroscopic properties, such as \(\beta\), can be influenced by introducing microscopic interactions between dimers and plaquettes.

The phase diagram of the close-packed model, as shown in Fig.~\ref{phasedia}, indicates that when \(\beta < \pi/2\), entropy fluctuations within the close-packed tiling patterns give rise to a fracton liquid phase. This phase is characterized by quasi-long-range order in the correlations between dimers and plaquettes.
\begin{align} 
&\langle E_z(0) E_z(r) \rangle =\frac{1}{\beta}\left(\frac{1}{z^2+x^2y^2}-\frac{2z^2}{(z^2+x^2y^2)^2}\right)
\end{align}
\begin{align} 
&\langle E_{xy}(0) E_{xy}(r) \rangle =\frac{1}{\beta}\left(\frac{1}{z^2+x^2y^2}-\frac{2x^2y^2}{(z^2+x^2y^2)^2}\right)\label{co}
\end{align}
Both exhibit power-law decay characterized by strong anisotropy. This observation leads to the conclusion that the extensive entropy associated with these close-packed patterns results in a liquid phase where dimer and plaquette configurations strongly fluctuate. Furthermore, the anisotropic power-law correlation gives rise to unique pinch point singularities in momentum space \cite{prem2018pinch,benton2021topological,nandkishore2021spectroscopic} with potential to be detected via neutron scattering.

The fracton liquid phase can transition into an ordered phase as the stiffness of the system increases. This transition from the liquid to the ordered phase is characterized by a fractonic version of the Kosterlitz-Thouless transition.
A particularly intriguing feature of these close-packed dimer-plaquette models is the phenomenon of \textit{UV-IR mixing}, where the low-energy effective theory is significantly influenced and dominated by discontinuities in field patterns\cite{seiberg2020exotic,lake2021rg}--operators with higher-order derivatives. This implies that high momentum modes can survive in the low energy spectrum and thus play a crucial role in driving the phase transition.
This characteristic implies that both the fracton liquid phase and its transition to order phase transcend the traditional renormalization group framework, as the critical low-energy behaviors are manipulated by local fluctuations at short wavelengths. Such phenomenon, denoted as UV-IR mixing\cite{xu2007bond,paramekanti2002ring,you2019emergent,you2020fracton,you2022plaq,han2022non} introduces a novel category of critical phenomena that extends beyond conventional renormalization paradigm.

\section{A general recipe for creating higher-rank fracton liquids} \label{sec:gen}

Just as exotic quantum materials can be scrutinized through topological quantum chemistry by analyzing and compiling a dictionary of quantum materials from the representations of crystalline groups, it would be enlightening to devise a concrete and universal recipe for evaluating and manipulating various fracton liquids in magnetic materials from the perspective of generalized symmetry and emergent higher-rank gauge theory\cite{sadoune2024human,yan2023classification,placke2023ising}. Beyond the practical utility of discovering novel fracton liquids in experimental settings, a crucial question on the agenda is to promote a coherent mapping between the internal gauge structure of higher-rank spin liquids and pinch-point singularities in the spin structure factors. This correspondence would enable us to discover and collect candidates for fracton liquid materials through the characteristic pinch-point neutron scattering data.

\subsection{General set up}
Ref.~\cite{benton2021topological} introduces an innovative approach to evaluating and manipulating emergent tensor gauge theories within the realm of frustrated magnetism. This method is based on linking the ground state constraints—determined by the specific spin Hamiltonian with local conservation laws—to the topological characteristics of a vector function in momentum space, denoted as \(\mathbf{L}(\mathbf{q})\).
Consider a general spin model whose low-energy space exhibits an emergent gauge symmetry. This symmetry imposes a local constraint on the Heisenberg spins within each local cluster, shaping them into a specific form as follows:
\begin{eqnarray}
\sum_{i \in c} \eta_i {\bf S}_i =0 \ \forall \ c
\label{eq:real_space_constraint}
\end{eqnarray}
In this context, \(c\) represents certain real-space clusters, and \(\eta_i\) denotes specific real coefficients. 
Such local constraint can be reformulated in momentum space as:
\begin{eqnarray}
&&\sum_{m=1}^{n_{u}}  L_m^{(p)} ({\bf q})^{\ast} {\bf S}_m({\bf q})   =0 \ \ \forall  \ \ {\bf q}, \ p 
\label{eq:q_space_constraint}
\\
&&L_m^{(p)}=\sum_{i \in m \in c_p} \eta_i \exp(i {\bf q} \cdot ({\bf r}_{c_p}-{\bf r}_i))
\label{eq:L_def}
\end{eqnarray}
Without loss of generality concerning the crystalline structure, we assume there are \(n_u\) sites per cluster cell, with the summation in Eq.~\ref{eq:q_space_constraint} extending over the cluster cells. The term \(\mathbf{S}_m(\mathbf{q})\) represents the lattice Fourier transform of the spin configuration on the \(m^{\text{th}}\) sublattice. The number of vectors \(\mathbf{L}^{(p)}\) corresponds to the number of local constraints, each leading to an emergent gauge symmetry. This conceptualization allows us to view each independent vector \(\mathbf{L}^{(p)}\) as representing a generalized Guass's law. Given that the low-energy Hilbert space is spanned by spin patterns satisfying the gauge symmetry in Eq.~\ref{eq:q_space_constraint}, we can interpret the spin correlation functions in terms of a projective matrix \(\mathcal{P} (\mathbf{q})\). By projecting out all vectors \(\mathbf{L}^{(p)}(\mathbf{q})\), the resulting correlation is defined within the gauge-invariant subspace, subject to Eq.~\ref{eq:real_space_constraint}.
\begin{eqnarray}
\langle {\bf S}_m(-{\bf q}) \cdot 
{\bf S}_n({\bf q}) \rangle= \frac{1}{\kappa} 
\mathcal{P}_{mn} ({\bf q})
\label{eq:Sq_proj}
\end{eqnarray}
The variables \(m\) and \(n\) are sublattice indices, and \(\kappa\) is a normalization constant.

In momentum space, when a vector \(\mathbf{L}^{(p)}(\mathbf{q})\) vanishes or becomes linearly dependent on others, it leads to singularities in the projection matrix \(\mathcal{P}_{mn}(\mathbf{q})\). These conditions indicate that the emergent gauge symmetry and the conservation law of spin multipole moments can uniquely induce momentum space singularities in the structure factor, akin to pinch points observed in tensor spin liquids\cite{prem2018pinch}. 

\begin{figure}[h!]
    \centering
   \includegraphics[width=0.8\textwidth]{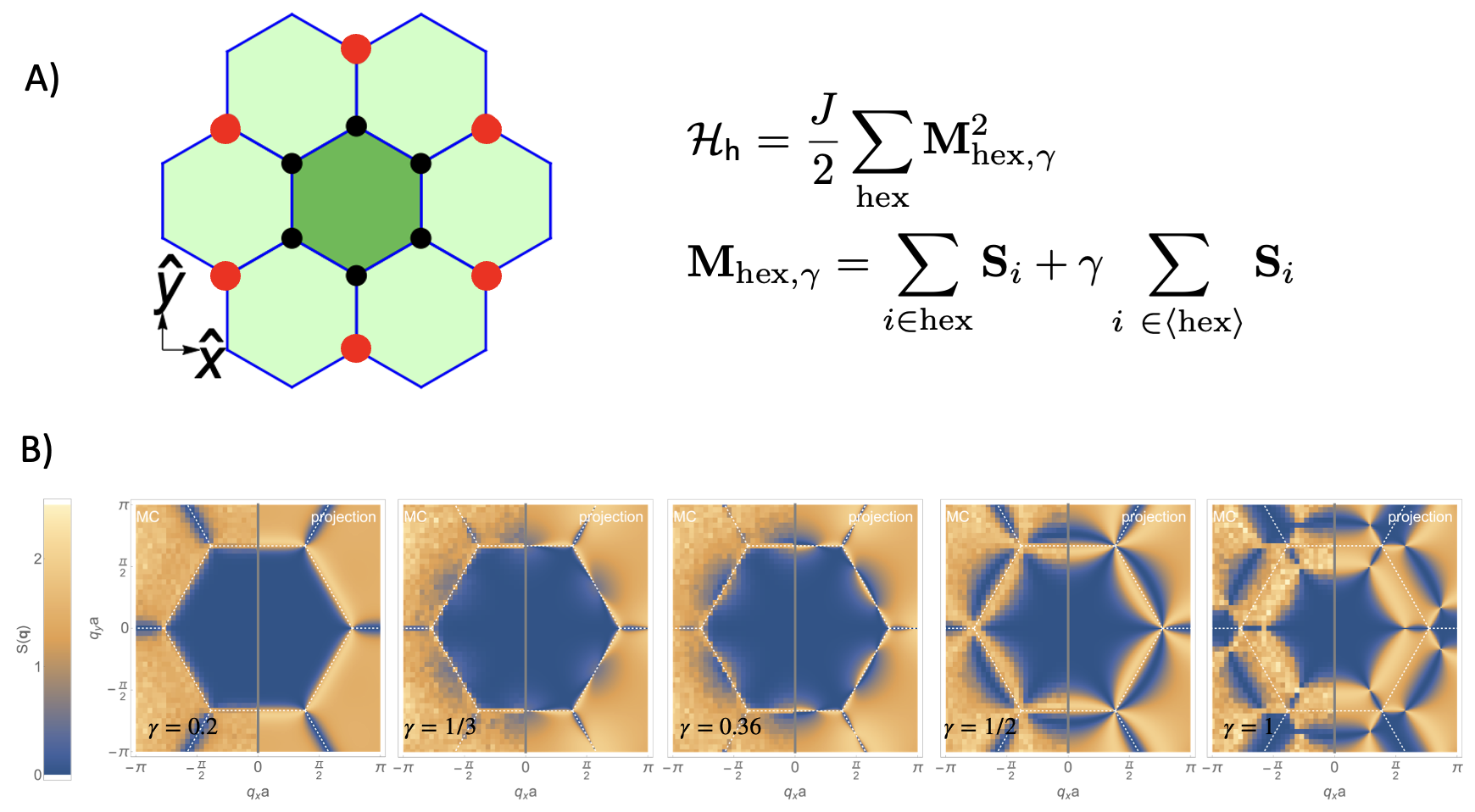}
    \caption{A) The spin model on hexagon lattice with spin cluster interaction. The Hamiltonian effectively pins the total spins in each cluster including six spins on the hexagon (dark) and the six spins (red) connected to the exterior of the hexagon with a relative ratio parameterized by $\gamma$.
    B)Evolution of $S({\bf q})$ across distinct higher-rank spin liquids. Part of the figure is reproduced in Ref.~\cite{benton2021topological}.}
    \label{fig:honeycomb}
\end{figure}

\subsection{Kick-off example: Engineering Fracton Liquids on a Honeycomb Lattice}

As an example, we review the spin Hamiltonian presented in Ref.~\cite{benton2021topological} that spreads a class of classical spin models on the honeycomb lattice:
\begin{eqnarray}
&&\mathcal{H}_{\sf h}=
\frac{J}{2} \sum_{{\rm hex}} {\bf M}_{{\rm hex}, \gamma}^2
\end{eqnarray}
\begin{eqnarray}
&&{\bf M}_{{\rm hex}, \gamma}=\sum_{i \in {\rm hex}} {\bf S}_i + \gamma \sum_{i \ \in \langle {\rm hex} \rangle} {\bf S}_i 
\label{eq:Mhex}.
\end{eqnarray}
The first term in Eq. (\ref{eq:Mhex}) represents the summation of spins forming a hexagon, while the second term encompasses spins linked to the hexagon's exterior, as depicted in Fig.~\ref{eq:Mhex}. This Hamiltonian constrains the total `effective spin number'\footnote{The effective spin number is defined by parameterizing the sum of the six spins located on the hexagon and the six additional spins connected to the hexagon's exterior, with this total adjusted by a modulation ratio, \(\gamma\).} within each \(\mathbf{M}_{\text{hex}}\) cluster, thereby facilitating a range of emergent gauge symmetries through adjustments of the dimensionless parameter \(\gamma\). The unique gauge structure associated with each \(\gamma\) value becomes apparent through observing distinct momentum-space singularities in the spin correlation, as shown in Fig.~\ref{fig:honeycomb}.
For small \(\gamma\) values, the formation of vortices in \(\mathbf{L}(\mathbf{q})\) is observed at the Brillouin Zone corners, along with the emergence of pinch points in \(S(\mathbf{q})\). At \(\gamma=1/3\), pairs of topological defects with opposite charges appear at the zone boundary and move towards the zone corners. When \(\gamma\) reaches \(1/2\), these defects merge into entities carrying a charge of \(Q=\pm2\), leading to the formation of 4-fold pinch points in the structure factor, indicating a classical rank-2 spin liquid phase in 2D. Upon further increasing $\gamma$, the defects separate, transitioning the system into a rank-3 spin liquid phase characterized by 6-fold pinch points.
 
Notably, the zeros of \(\mathbf{L}^{(p)}(\mathbf{q})\) represent a topological charge, originating from the `generalized Guass-law' that engenders the emergent gauge symmetry. As such, they remain stable against minor changes to the spin cluster interaction in Eq.~(\ref{eq:real_space_constraint}). More significant alterations might result in the creation or annihilation of additional topological defects, leading to a transition to a different spin liquid phase characterized by an increased or reduced number of singularities in the momentum space correlations. By modifying Eq.~\ref{eq:real_space_constraint}, one can also combine topological defects into entities with higher charges and develop a new category of higher-rank spin liquids, potentially facilitating phase transitions between them.

\subsection{Explore fracton liquids from synthetic quantum matter}
In recent studies\cite{yang2021z,chamon2021superconducting,zhou2021experimental}, it has been shown that a wider variety of spin models with local interactions can be meticulously engineered through the manipulation and control of superconducting circuits within the D-Wave DW-2000Q quantum annealer. These systems offer programmable couplings between pairs of spins (or superconducting qubits) and the ability to apply an external transverse field. 
Inspired by the general protocol of fracton liquid proposed\cite{benton2021topological}, one may expect that a vast range of tensor spin liquids can be designed and constructed using programmable spin clusters, as facilitated by this innovative protocol.

Another promising platform for realizing fracton liquids is proposed in Rydberg atom systems under tweezer arrays. The capacity for individual atom control, combined with strong interactions over considerable distances, renders these arrays an ideal environment for the manifestation of exotic many-body quantum states. 
Ref.~\cite{verresen2021prediction,ebadi2022quantum,weimer2011digital,giudice2022trimer,slagle2022quantum,semeghini2021probing} outlines a scalable method for preparing a wide variety of fracton states via Rydberg atom arrays, where the intrinsic atomic interactions are allowed to evolve over a finite period, followed by partial measurements. Remarkably, this protocol has predicted the ability to prepare 3D fracton states with high fidelity exceeding 0.99\cite{verresen2021efficiently}.
With recent experimental advancements in 3D Rydberg atom controls, this approach not only advances the experimental realization of fracton order but also offers a new pathway to explore quantum dynamics in fracton systems. 

In a complementary approach, Ref.~\cite{zhang2023synthetic} proposed the manipulation of synthetic higher-rank gauge fields as a method to generate fracton excitations. In this scheme, a rank-2 electric field is generated through the interplay between a strong linear potential\cite{scherg2021observing} and a subdued quadratic potential. Alternatively, the application of external lasers can facilitate photon-assisted tunneling in a linearly tilted lattice, thereby replicating similar rank-2 tensor gauge fields. Such rank-2 electric field in an optical lattice generates a new type of Bloch oscillations, characterized by the vibration of a dipole rather than the response of a single particle or hole, which is akin to the scalar charge theory in fracton liquids\cite{rasmussen2016stable}. 
This protocol also enables the establishment of the dipolar Harper-Hofstadter model and offers a pathway for extension to fermionic and multi-component systems.

An alternative pursuit involves exploring the analogy of fracton physics in complex systems, such as isostatic lattices, whose floppy modes and self-stress states exhibit dynamics akin to fractons with restricted motion\cite{mao2018maxwell}. While the floppy modes and self-stress states can be characterized by the index theorem as a local generalization of the Maxwell-Calladine counting rule, it would be fascinating to see how this local counting rule mirrors the generalized Gauss's law found in fracton literature.

\section{Outlook}

While the field of fracton liquids and tensor gauge theories has seen rapid growth, it remains in an early stage of development, with many open theoretical and experimental challenges. Significant advances have been made in understanding gapless fractons driven by symmetry breaking in charge multipole or subsystem symmetries\cite{stahl2022spontaneous,lake2021bose,anakru2023non,lake2023non}, apart from fracton liquids derived from high-rank gauge theories. 
Another intriguing area of exploration is whether a continuous phase transition exists between various fracton phases and how UV-IR mixing alters their critical behavior.
Additionally, given the increasing interest in open quantum systems and the effects of decoherence quantum channels on quantum field theory, exploring the fate of higher-rank gauge theories under decoherence remains a significant item on the agenda.

In this review, we have charted the recent theoretical development of fracton liquids. 
It is important to emphasize that significant progress has also been made in other areas of fracton research not included in this review and we recommend readers refer to Refs.\cite{pretko2020fracton, nandkishore2019fractons, gromov2024colloquium} for complementary reading. These areas include the field theory of gapped fracton topological orders, non-equilibrium thermodynamics in systems with subsystem or charge-modulated symmetries, hydrodynamics, and quantum information aspects of fracton.
The fracton frontier remains largely unexplored, and there remain no doubt new continents to discover. We look forward to new researchers joining in the exploration of this new frontier and advancing our understanding of the field.

\end{document}